\documentstyle[12pt]{article}

\newcommand{\ue}%
{\mbox{c\hspace{-0.4em}\rule[ 0.5ex]{0.3em}{0.04ex}\hspace{0.1em}}}
\newcommand{\Ue}%
{\mbox{C\hspace{-0.6em}\rule[ 0.75ex]{0.4em}{0.04ex}\hspace{0.2em}}}
\newlength{\signlength}%
\newcommand{\gs}%
{\settowidth{\signlength}{$<$}%
\raisebox{-0.35\signlength}%
{\makebox[1.7\signlength][c]{\makebox[0.pt]{$\sim$}%
\raisebox{0.6\signlength}{\makebox[0.pt]{$>$}}}}}
\newcommand{\ls}%
{\settowidth{\signlength}{$<$}%
\raisebox{-0.35\signlength}%
{\makebox[1.7\signlength][c]{\makebox[0.pt]{$\sim$}%
\raisebox{0.6\signlength}{\makebox[0.pt]{$<$}}}}}
\renewcommand{\ge}%
{\settowidth{\signlength}{$>$}%
\setlength{\unitlength}{0.1\signlength}%
\mbox{\begin{picture}(17,7.5)%
\linethickness{0.045\signlength}%
\put(0.0,1){\makebox(17,7.5){$>$}}%
\multiput(4.5,-1)(0.25,0.125){32}{\line(1,0){0.25}}%
\end{picture}}}
\renewcommand{\le}%
{\settowidth{\signlength}{$>$}%
\setlength{\unitlength}{0.1\signlength}%
\mbox{\begin{picture}(17,7.5)%
\linethickness{0.045\signlength}%
\put(0.0,1){\makebox(17,7.5){$<$}}%
\multiput(12.,-1)(-0.25,0.125){32}{\line(-1,0){0.25}}%
\end{picture}}}
\title{Dynamical Dimension Reduction in Underdoped High Temperature
Superconductivity}
\author{Sergeeva G.G. $^{1}$, Vakula V.L.$^{2}$}
\date{}
\begin{document}
\maketitle
\thispagestyle{plain}

\begin{center}
$^{1}$ National Science Center "Kharkov Institute of Physics
and Technology", Academicheskaya st. 1, 61108, Kharkov, Ukraine

$^{2}$ B.Verkin Institute for Low Temperature Physics and Engineering
of the National Academy of Sciences of Ukraine, 47 Lenin Ave., Kharkov
61103, Ukraine
\end{center}

\begin{abstract}
We discusse the supposition that both the pseudogap (PG)
and the superconducting (SC) states of underdoped high-$T_c$
superconductors (HTSC) result from dynamical dimension
reduction when HTSC behave on cooling if their dimensionality is
changed. It is shown that the transition to the PG
state occurs at the temperature $T^{\ast}$ as a dimensional crossover
when charge motion changes from three-dimensional (3D) to
two-dimensional (2D).  Namely two-dimensionality at $T^{\ast}>T>T_c$
is responsible for the crucial role of Jahn-Teller (JT) distortions
which bound up with holes and form
delocalized JT polarons and localized the three spin polarons in
copper-oxygen planes. This leads to charge ordering in copper-oxygen
planes and removes the competition
between pairing of carriers and their localization on the JT distortions.
As the temperature is lowered below
$T_{cr}<T^{\ast}$, the local "hole- JT polaron" pairs (i.e.
zero dimensional 0D SC fluctuations) are generated in $CuO$ planes.
At $T_{cr}>T>T_c$ the SC transition occurs as a sequence of two
crossovers for SC fluctuations: 0D$\rightarrow$2D crossover
and then 2D$\rightarrow$3D crossover. Some experimental
evidence of the local "hole- JT polaron" pairing and some
results of the study of dynamical dimension reduction in the
PG and SC states are discussed.

\end{abstract}

{\large
\section {Introduction}}

Despite of the intensive research of the nature of high temperature
(HT) superconductivity the questions about the pairing mechanism and
about the nature and number of carriers remain open. Today for the
normal state at $T>T^{\ast}$ for underdoped (UD) high temperature
superconductors (UD HTSC) the
two-component model of carriers can be considered as a firmly
established fact: small polarons and holes are the heavy and light
carriers respectively. Here $T^{\ast}$ is the temperature of the
transition to the pseudogap (PG) state in which $CuO$ plane are in
stripe state. For normal state the measurements of optical conductivity
for UD HTSC \cite{Orenstein}- \cite{Bi} provided
the first evidence for coexistence of these two carrier types.
Later from spin susceptibility measurements
for $La_{2-x}Sr_x Cu O_4$ the doping dependence of the part
for each carrier type was determined \cite{Muller}, but the character of
the polaron state (it is the polaron or bipolaron) was unclear.
This question for cuprate HTSC is of fundamental importance because
the existence of Jahn-Teller (JT) small polarons for doped
antiferromagnets (AF) with large value of dielectric constant
and mobile light oxygen ions \cite{Hock} namely
was the initial point for HTSC searches.  Upon doping
the extra holes are localized on the transition metal ions that
causes the change of their valence and leads to the strong
JT distortions. At the doping increase the transition of AF
into metal occurs with two carrier types (for example in $W O_{3-x}$
\cite{Salje}), and the superconducting (SC) transition with enough
high value of $T_c$ is in principle possible.  Really the surface
superconductivity with
$T_c =90K$ recently was observed for $W O_3$ doped by $Na$ ions
\cite{Reich}. 

Now it is clear that understanding of the PG state nature at                                             
$T^{\ast}>T>T_c$ will provide a clue to HT superconductivity.
This supposition is based on the following facts: 1) the change
of the density of states starts at $T\sim T^{\ast} $ and continues
on up to $T_c$; 2) at $T_c$ the coherent SC state is formed with
practicably no effect on
the density of states; 3) there is a lot of evidence
of an evolution of the SC fluctuations at $T^{\ast}>T>T_c$ (see Refs.
in \cite{Mihailovic}). In Ref.\cite{Mihailovic} under Bose-Einstein
Condensation theory the three component model of the PG state was
proposed in which bipolarons were the third component of carriers.

In this paper the authors discussed a supposition that the PG and SC
states in UD HTSC result from dynamical dimension reduction when HTSC
behaves as if its dimensionality changes at the lowering of the
temperature  below $T^{\ast}$.
At that the transition to the PG state is the dimensional 3D
$\rightarrow$ 2D crossover of charge motion. At $T<T^{\ast}$ namely
two-dimensionality leads to the crucial role of JT distortions in $Cu O$
planes which bound up with holes and form delocalized JT polarons
and the localized three spin polarons in $Cu O$ planes. The chains
of the latter form in $Cu O$ planes a narrow stripes with distorted
low temperature tetragonal-like lattice.
This means that the dimensional crossover at $T=T^{\ast}$ leads to the
charge ordering in $Cu O$ planes, and removes the competition between
pairing of carriers and their localization on the JT distortions.
At $T_{cr}<T^{\ast}$ the "hole-JT polaron"
pairing occurs which are zero-dimensional (0D) SC fluctuations.
At $T_{cr}>T>T_c $  a sequence of two crossovers of the SC
fluctuations occurs: the first is the crossover 0D$\rightarrow $2D,
and second is the crossover  2D$\rightarrow $3D that leads to three-
dimensional SC transition.

{\large
\section {The transition to the pseudogap state  as
3D$\rightarrow $2D crossover of the charges motion}}

For UD HTSC incoherent interlayer tunnelling the charge transfer
along $c$ axis is the result of the thermal fluctuations at
\begin{equation}\label{1}
k_B T>t^{2}_c(T)/t_{ab}.
\end{equation}
Here $t_c$ and $t_{ab}$ are the interlayer hopping rates of the charges,
$k_B$ is the Boltzman constant. At the temperature decreasing
\begin{equation}\label{2}
k_B T \simeq  t^{2}_c(T)/t_{ab}
\end{equation}
thermal fluctuations limit out the interlayer tunnelling. This
leads to the dimensional crossover  3D$\rightarrow $2D of charges
motion when at the temperature
\begin{equation}\label{3}
k_B T^{\ast}= t^{2}_c (T^{\ast})/t_{ab}
\end{equation}
the charges in $Cu O$ plane moves as two-dimensional ones. This
means that the transition to the PG state is the result of
dynamical dimensional reduction.

Lowering of dimensionality leads to the crucial role of any
disorder and to the changes of character of the SC fluctuations.
One in the first attempts to consider in a self-consistent way
the competition between pairing of the carriers and their
localization on the defects were made in works \cite{Loktev}-\cite{Loktev1}.
At $T<T^{\ast}$ for underdoped cuprate HTSC two-dimensionality
leads to the crucial role  of JT distortions
around two adjacent $Cu ^{2+}$ ions which bound up holes
\cite{Lifshits} -\cite{Nagaev}, and
form the quasilocal states ( delocalized JT polarons) and local
states (the localized
three spin polarons \cite{Kochel}, or ferrons \cite{Nagaev}).

Mobile delocalized JT polaron is the quasilocal state of hole bounded
up with complex of two adjacent distorted by JT interactions "squares"
$Cu^{2+}+4O^{2-}$ with common oxygen ion, and $Q_2$ phonon normal mode
leads to the oscillations of "squares" (see Fig. 1a). Total spin of
JT polaron is equal 1/2, and spins of two $Cu^{2+}$ ions are antiparallel.
In $Cu O$ plane these JT polarons form wide $U$ stripes with nearly
undistorted low temperature orthorhombic-like lattice \cite{Bianc}.

{\large
\section {The localized three spin polarons}}

The studying ferromagnetic self-trapped states
of a charge carriers in a doped AF crystal was began in
1968 by Nagaev (see Refs.in \cite{Nagaev}). Later this type of states
has been proposed by Emery and Reiter, and at first it was
observed and named "the three spin polaron" by Kochelaev et al.
(\cite{Kochel}, and Refs. there).
Electron-paramagnetic resonance (EPR) measurements provide
experimental evidence of the existence of the three spin polarons
and the presence of dynamical JT distortions with normal modes
$Q_4$ and $Q_5$, which have tetragonal symmetry
and lead to exchange spin-phonon interaction similar to the
Dzyaloshinskii-Morya
interaction \cite{Kochel2}
\begin{equation}\label{4}
H_{s-ph}=\frac{6\lambda G J^{2} }{a\Delta} \sum\limits_{k,q}^{}[cos (ak_x)+
cos(ak_y)]\cdot \exp(izq')[(S^{y}_k S^{z}_{k-q} -
\end{equation}

\begin{center}
$$-S^{y}_{-k-q} S^{z}_k) Q_{4k}+ (S^{x}_k S^{z}_{-k-q}-S^{x}_{k-q}
S^{z}_k)Q_{5q}],$$
\end{center}

where $J$ is the exchange antiferrognetic coupling constant, $\lambda$ is
spin-orbit coupling constant, $\Delta$ is an average splitting between
energy levels, $S^{x}_k$ is a two dimensional Fourier's transforms a
component of the spin operator, $q$ and $q'$ are projections of three
dimensional wave vector on the $Cu O$ plane and $c$-axis respectively,
and $G$ is electron-phonon coupling constant. This small interaction
cannot pin the three spin polaron in 3D system but at $T<T^{\ast}$ for
2D system local state exists at any values of the interactions in the
frame of I.M.Lifshits theory \cite{Lifshits}.

The isotope effect on the EPR linewidth (it doubles) upon
substitution $ ^{16}O\rightarrow ^{18}O$ quantitatively adjusts with
exchange spin-phonon interaction assistance in (4), and can be indirect
evidence of 2D nature of the PG state: $T^{\ast}|_{O^{16}}$=110 K, and
$T^{\ast}|_{O^{18}}$=180 K (\cite {Muller1} and Refs. there).
   The three spin polaron with parallel spins of two  adjacent
$Cu^{2+}$ ions and with total spin 1/2 is
localized  state of a hole bounded up with two distorted "squares"
 $Cu^{2+}+O^{2-} $ with common oxygen ion (see Fig.1b). Their chains
form in $Cu O$ plane narrow stripes with distorted low
temperature tetragonal-like  lattice (D stripes \cite {Serg}-\cite {Bianc}).
Thus in spite of localization of the carriers part on dynamical JT
distortions with normal modes $Q_4$ and $Q_5$, charge ordering in
$Cu O$ plane removes the competition between pairing of carriers and
their localization on the JT distortions.

{\large
\section {The possibility of the JT polaron and hole pairing}}

For UD HTSC the coexisting at $T>T^{\ast}$ of polarons and holes
stimulated the interest to the studying the possibility of their
pairing, but at that the mechanism of the suppression of on-site
Coulomb repulsion $U_c$ for two particles is the main problem for
HTSC. In Refs.\cite{Barab}-\cite{Kudin} the possibility of
such pairing was shown.  Kudinov \cite{Kudin} at the first shows
the principal possibility of on-site attraction of polaron and hole.
His conclusion is based on the following results : (i) polarons
lead to the band narrowing and to the polaron shift of the energy
$E_p=g^{2}_{JT}/ 2M\omega^{2}$ (here $g_{JT} $ is the constant of
JT interactions of holes with mobile oxygen ions, M is the effective
mass of JT mode); (ii) at $(-E_p +U_c)<0$ both as the compensation
of Coulomb repulsion, and so the on-site attraction between the hole
and polaron take place.

In Refs. \cite{Barab}-\cite{Kudin} the Zhang-Rice polarons
\cite{Zhang} were taken into account which have total spin equal
zero, and their pairing with holes cannot lead to the local
pairs. But Kudinov model easily generalizes for JT polarons with
spin equal 1/2 (see Fig. 1a), if after canonical Holstein-Lang-Firsov
transformations $U=\prod\limits_{m}^{}\exp(ix_0 \sum\limits_{\sigma}^{}
n_{m,\sigma} p_m)$, in Hamiltonian all two-particle renormalizated
interactions are taking into account (between holes, between JT
polarons, as well between the hole and JT polaron):
\begin{equation}\label{5}
V=\sum\limits_{m,g,\sigma}^{}J(g)|b^{+}_{m,\sigma}b_{m+g,\sigma}
+a^{+}_{m,\sigma}a_{m+g,\sigma}\cdot
exp[ix_0 (p_m -p_{m+g}] +a^{+}_{m,\sigma}b_{m+g,\sigma}|.
\end{equation}

Here $J(g)$ is non-renormalizated interaction constant between
holes, $x_0 =  g_{JT}/\hbar k_{JT}$, $k_{JT}$ is elastic constant of
JT normal mode, operators $a^{+}_{m,\sigma}$, and $b^{+}_{m,\sigma}=
a^{+}_{m,\sigma} exp (ix_0 p_m) $ create the hole and JT polaron on
site $m$, respectively, and $p_m$ and $x_0$ are the impulse and
equilibrium coordinate of the common oxygen ion for two adjacent
"squares" $Cu^{2+}+O^{2-}$. One can see that strong JT interactions
lead to different renormalization of all two-particle interactions.
At $<\exp(ix_0 p_m)> \simeq \exp (-E_p/ \hbar\omega)$ \cite{Kudin}
generalized Kudinov's Habbarg Hamiltonian is equal
\begin{equation}\label{6}
H_H= \sum\limits_{m,g,\sigma}^{}(2(-E_p +U_c)
  n_{m,{\uparrow}} n_{m,{ \downarrow}} +
J^{\ast}(g) a^{+}_{m,\sigma}b_{m+g,\sigma}),
\end{equation}
where $ J^{\ast}(g)= J(g) exp (-E_p/ \hbar \omega)$. All many
particles interactions are exponential renormalizated that leads to the
small contribution of bipolarons in the energy relatively with the
contribution of the on-site "hole- JT polaron" pairs (because former
contribution has exponential dependence on the distance between two
JT polarons \cite{Kudin}). As it shown by Kudinov in BCS model the
pairing hole and JT polaron leads to the possibility of the SC transition
in $CuO$ plane with the temperature $T_{cr} \sim|E_p - U_c|$.  At that
the local pair "hole-JT polaron" occupies the of JT polaron complex,
and has coherent length $\xi_{ab} \sim4R_{Cu-O}$ ($R_{Cu-O}$ is the mean
distance between $Cu$ and $O$ ions in $CuO$ plane).

Thus, the dimensional crossover at $T^{\ast}$ leads to the
charge ordering in $CuO$ plane and removes the competition between
pairing of the carriers and their localization on the JT distortions.
For cuprate HTSC the coexistence at $T<T^{\ast}$ of holes and JT
polarons is fundamentally important but decisive role belong to
the latter. JT polarons lead to polaron shift of the energy,
and to the compensation of on-site Coulomb repulsion between
JT polaron and hole in $Cu O$ planes. This leads at $T_{cr}<T^{\ast}$
to the local pairing of JT polarons and holes, i.e. to zero-dimensional
(0D) SC fluctuations \cite{Kulik}- \cite{Solov}. The temperature
lowering leads to the increasing of the coherent length $\xi_{ab}(T)$,
so that at enough big $\xi_{ab}$ local pairs begin overlap, and at
$T_{2D}<T_{cr}$ the dimensional crossovers 0D$\rightarrow$2D of the SC
fluctuations occurs. Second crossover 2D$\rightarrow$3D of the SC
fluctuations occurs at $T_{3D}<T_{2D}$ \cite{Serg2}- \cite{Serg3}.

 {\large
\section { Dynamical dimension reduction for the PG and SC states of
UD HTSC}}

The transition to the 2D SC fluctuations with the temperature dependence
of the coherent length $\xi_{ab}(T)=\xi_{ab}(T_{BKT})(T/T_{BKT}-1)^{-1/2}$
leads to the semiconducting dependence of the $c$-axis resistivity with
the probability of the charge transfer which is depending on the
temperature \cite {Serg1}, \cite {Serg3}:
\begin{equation}\label{7}
t_c(T)=\frac{\xi^{2}_c}{\xi^{2}_{ab}}(\frac{T}{T_{BKT}}-1),
\end{equation}

where $T_{BKT}$ is Berezinskii-Kostelits-Thouless temperature (BKT) of
the 2D SC transition for the isolated $CuO$ plane, and $\xi_c$ and
$\xi_{ab}$ are the values of the coherent lengths at $T=T_{BKT}$. At
sufficiently small $t_c (T)$ the Kats inequality \cite{Kats}
$T_c /E_F \ge t_c(T_c)$ determines the temperature of the SC transition
which occurs as two dimensional one with small region of the 3D SC
fluctuations (here $E_F$ is Fermi energy)
\begin{equation}\label{8}
T_{BKT}< T_c < T_{BKT}(1-\frac{\xi^{2}_{ab} T_{BKT}}{\xi^{2}_c E_F
-\xi^{2}_{ab}T_{BKT}})
\end{equation}
For example, from the analysis of the resistivity
measurements in single crystal Bi-2212 with $T_c = 80K$ \cite {Serg1}
it follows that the region of the (0D+2D) SC fluctuations
$(T_{cr} - T_{3D})\sim120 K$, and the region of the 3D SC fluctuations
$(T_{3D} - T_c)\sim 10 K$, and $T_{BKT} \sim 0.7 T_c \sim 56K$.

At the analysis of the SC state in \cite{Serg4} it was shown that at
$T<0.7T_c$  the coherence length $\xi_c(T)$ becomes less than the
interlayer distance. This means that once more dynamical dimension
reduction occurs at which the 3D SC state changes into the 2D SC
state. The boundary of the region of "three-dimensionality" of
the SC state can be determined as the temperature at which the
two universal temperature dependencies of the ratio of the squares
of the penetration depths of a magnetic field directed along $c$
axis,$\lambda^{2}(0)/ \lambda^{2}(T/T_c) $, are crossing: one is
determined by the 3D SC fluctuations in the BCS theory
\begin{equation}\label{9}
\lambda{2} _1(0)/ \lambda {2} _1(T/T_c)=2(1-T/T_c),
\end{equation}
and another is universal dependence for the 2D degenerate system
\begin{equation}\label{10}
\lambda^{2}_2 (0) / \lambda^{2}_2 (T/T_c) = \exp \frac {T e^{-1}
\lambda^{2}_2(T/T_c)}{T_c \lambda^{2} _2 (0)}.
\end{equation}

The temperature interval of "three-dimensionality" of the SC state
occupies only two temperature regions,
$\Delta_1=T_c - T_{BKT}$ near $T_c$, and $\Delta_2 = T_{BKT}-T_g$
where $T_g$ is the temperature of the transition into spin cluster
glass SC state (state (2+3) on fig.2)\cite{Nieder}. For example,
$\Delta_1 \sim11K$ for $La_{1.85} Sr_{0.15} Cu O_4$ , and region of
"three-dimensionality" of the SC and PG states near  $T_c$ is equal
$T_{3D}-T_{BKT} \sim15.5K$ \cite{Serg4}. For UD HTSC last dimensional
reduction  occurs at temperature $T_g<<T_{BKT}$ when the  2D SC state
changes on 3D spin cluster glass SC state \cite{Nieder}, and the values
of $T_g< 20 K$.

Thus, for UD HTSC the SC transition has two-dimensional character
(according to the Kats definition \cite{Kats}) with limited total
region of "three-dimensionality" of the SC and PG states. The
peculiarities of normal, the PG and SC states can be understood
taking into account the effect of dynamical dimension reduction
under cooling when HTSC behaves as if its dimensionality repeatedly
changes at $T^{\ast}, T_{cr}, T_{2D}, T_{3D}, T_{BKT}, T_{g}$ (Fig.2).
At that the PG transition is a transition to 2D carriers motion at
$T^{\ast}$, and the SC transition occurs as a sequence of dimensional
crossovers of the SC fluctuations.

{\large \section{ Discussion}}

For UD HTSC at dimensional crossovers of the SC fluctuations
holes number $n_h$ and polaron number $n_p > n_h$ \cite {Muller},
are decreasing that is according with the observation of the
noticeably change of the density of states at $T^{\ast}>T>T_c$
\cite{Mihailovic}. For example, at the Hall effect measurements for
$YBa_2, Cu_3 O_{6+x} (T_c = 87.4K)$ was found out that $n_h$ decreases
at lowering the temperature from $240K$ up to $100K$ \cite{Solov}:
$n_h (240K)\sim5.4\cdot10^{21}cm^{-3};
n_h (100K)\sim2.7\cdot10^{21}cm^{-3}$.

For HT superconductivity the conclusion about the decisive role of
"hole-JT polaron" pairing qualitatively comes to an agreement with the
doping dependence of the part of each carriers type relatively of
total carriers number: in Ref. \cite {Muller} for $La_{2-x}Sr_x Cu O_4$
it was shown that the value of $T_c \rightarrow T_{c,max}$ (at"hole-JT
polaron" pairing $T_c \sim n_h\cdot n_p$) at the hole concentration
$\sim0.15$ on ion $Cu^{2+}$ when  $\frac{n_p}{n_p+n_h} \sim 0.6$ and
$\frac{n_h}{n_p+n_h} \sim 0.4$.

The temperature dependence
of the resistivity also evidences about the coexistence of carriers
and local pairs "hole-JT polaron" in the PG state \cite{Kulik}-
\cite{Bales}, \cite{Solov}.The studying fluctuational in-plane
conductivity \cite{Solov} shown that the interactions of fluctuational
pairs with the carriers are weak, and the contributions into
conductivity of the 0D, 2D and 3D SC fluctuations are been identified.

The measurements of optical conductivity are the convincing evidence
of "hole- JT polaron" pairing where it is shown that at $T<T^{\ast}$
the $c$-axis component of electronic kinetic energy and carriers mass
double \cite{Basov}. Once more convincing example of the coexistence of
the holes, polarons a local pairs "hole-JT polaron" in the PG and SC
states is the observation of a doublet structure of two-magnon absorption
band with maxima at $2.15 eV$ and $2.28 eV$ in two  metallic films
$YBa_2, Cu_3 O_{6+x}$ (x=0.5 and x=0.85 ) (see fig.3) \cite {Eremen}.
Its first component with energy $\omega\approx \Delta_{CT} + 3J$ is
identical to that which was observed in doped AF with $x=0.3$, and was
caused by two-magnon absorption at interband transition of polaron
(here $J\sim0.13 eV$ is the exchange energy, $\Delta_{CT}$ in transfer
energy). It is seen that the doublet structure appears in the PG state,
and becomes more prominenet  in the SC state.

It is known that in metal phase of the sample at $T>T^{\ast}$ this
component is not observed. For metal films at $T< T^{\ast}$ the
observation of this component together with second component with
$\omega \approx \Delta_{CT} + 4J$ undoubtedly attest to the fact
that developed antiferromagnetic  fluctuations exist in the PG
and SC states \cite{Eremen}. We conjecture that this doublet was
caused by the fulfillments of the condition of the "triple resonance"
for two-magnon absorption, similar to that at Raman scattering for
undoped AF \cite{Chub}-\cite{Morr}.  JT polaron absorbs of photon
(with energy $\omega$) and transfer into valence band. Two transfers
of the charge with energy $t$ (between $Cu^{2+}$ and oxygen ion within
JT polaron complex, there and back) lead to the radiation of two
magnons with the frequencies $\Omega_q , \Omega_{-q}$ at the resonance
condition
\begin{equation}\label{11}
\omega \approx \Delta_{CT} + 2t + \Omega_q + \Omega_{-q}.
\end{equation}
For UD HTSC at $T>T^{\ast}$ this condition is not practicable with
taken into account the interactions between holes and polarons, and
two-magnon absorption leads only to an essential asymmetry and big
width of right hand of two-magnon absorption band: for $YBa_2 Cu_3
O_{6+0.1}$ up to $\omega\sim3 eV$ \cite{Basov}-\cite{Morr}.  At
$T<T_{cr}$ holes and part of JT polarons $(n^{*}_p \sim n_h)$ are in
pairing state, and for unpaired  part of JT polarons $(n_p-n^{*}_p)$
resonance condition (11) takes place in both the PG and SC states.

The observation of  the doublet structure of two-magnon absorption band
evidences about (i) the existence of the JT polarons in PG and SC states,
and about (ii) essential charge heterogeneity of the SC state (the same
as for the PG state). The doublet structure of two-magnon absorption band
for the SC state is an indirect evidence of $d$-wave symmetry of the SC
order parameter for UD HTSC as well: at $T_c > T$ some part of unpaired
polaron $(n_p - n^{*}_p)$ percolates through the direction of wave
vector where the order parameter is equal zero.

Recent measurements of the $c-$axis charge pseudogap dependence on
the temperature and on the magnetic field H// $c$-axis \cite{Krasn}
indicate that the pseudogap neither disappears nor continuously transforms
into the SC gap below $T_c$, and this comes an agreement with our statement
that the PG transition occurs at $T^{\ast}$ as the dimensional
crossover from three-dimensional charges motion to the two-dimensional
one. The existence of the pseudogap at $T<T_c$ means that there are a
part of unpaired carriers which in the SC state moves only in $Cu O$
plane, and we above see that JT polarons are these unpaired carriers.

{\large
\section { Concluding remarks}}

For verification of various HTSC scenarios it is very important to
know the right answer on the question: is  the PG state at the
temperature $T^{\ast}>T>T_c$ a precursor of the SC state or not.
The problems of the decision of this question is connected with
non-three-dimensional character of the SC fluctuations for the
part of fluctuational interval $T_{cr}-T_{3D}$ where pairing
amplitude is non-zero, but phase rigidity is lost. This means that
at $T_{cr}>T>T_{3D}$ the standard measurements which are depending
on phase rigidity (such as Andreev reflection or insufficiently
high magnetic field) cannot be sensitive to the  SC fluctuations
with non-zero pairing amplitude, but without phase rigidity. For
example, the studying of the dependence of the fluctuations at
$T_{cr}>T>T_{3D}$ under sufficiently high (up to 33 T) magnetic
field H// $c$-axis can give the decisive answer on this question.

Our results mean that the temperature $T^{\ast}$ will be depending
on magnetic field $ H>H_{cr}$ which is parallel to $Cu O$ plane
(here $H_{cr}$ is the magnetic field which leads to suppression
of the 3D SC fluctuations \cite{Glazm}). At that $T^{\ast}$, and
all the temperatures of dimensional crossovers of the SC fluctuations
($T_{cr}, T_{2D}, T_{3D}$, and the temperature $T_{BKT}$) receive
the positive additions.

To summarize, in this paper for UD HTSC it is shown that the
coexistence of holes and Jahn-Teller polarons is fundamentally
important but decisive role belong to the latter. At that the
pseudogap transition occurs at $T^{\ast}$ as the dimensional crossover
from three-dimensional charges motion to two-dimensional.
Two-dimensionality leads to the charge ordering in $CuO$ plane and
removes
the competition between pairing of the carriers and their localization
on the Jahn-Teller distortions. At the cooling dimensionality repeatedly
changes in the pseudogap and in the superconducting state.  The
measurements under high magnetic field which is parallel to
copper-oxygen plane can give the decisive answer on the question:
is the pseudogap state a precursor of the superconducting state or not.

The authors are grateful to Profs. V.V.Eremenko and V.M.Loktev,
and Dr. V.N.Sa-\\
movarov for many comments and insightful discussions,
and to Dr. O.Dudko for information support.

\begin{center}
{\large Figure captions }
\end{center}

Fig. 1a). Quasilocal state of hole at $T<T^{\ast}$ (Jahn-Teller
polaron). Fig. 1b). Local state of hole at $T<T^{\ast}$ (the three spin
polaron). Light circles denote oxygen ions, dark circles denote
copper ions; small circles denote bound holes.

Fig. 2. Magnetic phase diagram for doped antiferromagnets
and underdoped HTSC: $T_N (\delta)$ is the doping dependence of Neel
temperature (AF state 1); $T_f(\delta)$ is the doping dependence of
the temperature of the ordering holes spin in $CuO$ plane (states 2, 5);
$T_g(\delta)$ is the doping dependence of the temperature of the
transition into spin cluster glass state (states 2 and (2+3));
$T_{2D XY}(\delta)$ is the doping dependence of the temperature of 2D XY
magnetic ordering for doped AF (state 4); $T_{BKT}(\delta)$ is the doping
dependence of the temperature of BKT transition for the SC state;
$T^{\ast}(\delta)$ is the doping dependence of the temperature
of the transition in the PG state; $T_c (\delta)$ is the doping
dependence of the temperature of the transition into the SC state
(3). The region of the PG state (5) limited by the  curves
$T_f, T_g, T_c$ and $T^{\ast}$.

Fig.3. Two-magnon absorption band ($E_0=2.15 eV$) for
films $Y Ba_2 Cu_3 O_{6+x}$: an AF dielectric film (x=0.3), and for two
metallic films (x=0.5, x=0.85). In the dielectric case the band is a
single peak centered at 2.15 eV. In metallic case it is a doublet
feature with maxima at 2.25 eV and 2.28 eV. The doublet structure
appears in the PG state and becomes more prominent in the SC state.

\end{document}